\documentstyle[12pt,aasms4]{article}

\begin{document}

\title{Probing $r$-Process Production of Nuclei Beyond $^{209}$Bi 
with Gamma Rays}
\author{Y.-Z. Qian\altaffilmark{1}, P. Vogel\altaffilmark{2}, 
and G. J. Wasserburg\altaffilmark{3}}
\altaffiltext{1}{T-5, MS B283, Theoretical Division, Los Alamos National
Laboratory, Los Alamos, NM 87545; qian@paths.lanl.gov.}
\altaffiltext{2}{Department of Physics, California Institute of Technology,
Pasadena, CA 91125; vogel@lamppost.caltech.edu.}
\altaffiltext{3}{The Lunatic Asylum,
Division of Geological and Planetary Sciences, California
Institute of Technology, Pasadena, CA 91125.}

\begin{abstract}

We estimate gamma-ray fluxes due to the decay of nuclei beyond $^{209}$Bi
from a supernova or a supernova remnant assuming that the $r$-process
occurs in supernovae. We find that a detector with a sensitivity of
$\sim 10^{-7}\ \gamma\ {\rm cm}^{-2}\ {\rm s}^{-1}$ at energies of
$\sim 40$~keV to $\sim 3$~MeV may detect fluxes due to the decay of
$^{226}$Ra, $^{229}$Th, $^{241}$Am, $^{243}$Am, $^{249}$Cf, and $^{251}$Cf
in the newly discovered supernova remnant near Vela. In addition, such a 
detector may detect fluxes due to the decay of $^{227}$Ac and $^{228}$Ra
produced in a future supernova at a distance of $\sim 1$~kpc. As nuclei
with mass numbers $A>209$ 
are produced solely by the $r$-process, such detections
are the best proof for a supernova $r$-process site. Further, they provide
the most direct information on yields of progenitor nuclei with $A>209$
at $r$-process freeze-out. Finally, detection of fluxes due to the decay
of $r$-process nuclei over a range of masses from a supernova or a supernova
remnant provides the opportunity to compare yields in a single supernova 
event with the solar $r$-process abundance pattern.

\end{abstract}


\section{Introduction}

Essentially all of the naturally occurring nuclei above the iron group
can be accounted for by nucleosynthesis via neutron capture 
(Burbidge et al. 1957; Cameron 1957). Most of these nuclei receive 
contributions from both the slow ($s$-) and the rapid ($r$-) neutron
capture processes. The measure of ``slow'' or ``rapid'' is defined by
comparison with the decay of unstable nuclei encountered in these
processes. Consequently, the $s$-process proceeds along or close to
the line of $\beta$-stability. By contrast, the $r$-process initially 
produces extremely neutron-rich progenitor nuclei far from stability,
which quickly decay to stable or long-lived nuclei after $r$-process
freeze-out.
As the nuclei immediately beyond $^{209}$Bi,
the heaviest stable isotope, have very short lifetimes, the nuclei with
mass numbers $A>209$, including the long-lived actinides (e.g., 
$^{232}$Th, $^{235}$U, $^{238}$U, and $^{244}$Pu), cannot be produced by the 
$s$-process. Therefore, the existence of the actinides demonstrates the 
operation of an $r$-process in nature. However, the astrophysical site
of the $r$-process remains to be established.

In this paper, we assume that the $r$-process occurs in core-collapse
supernovae (hereafter simply referred to as supernovae) and discuss
gamma-ray signals associated with the decay of nuclei beyond $^{209}$Bi
from a supernova or a supernova remnant. As these nuclei are produced
solely by the $r$-process, detection of such signals constitutes the best
proof for a supernova $r$-process site. Further, these signals are the most
direct probes of yields for progenitor nuclei with $A>209$ at $r$-process
freeze-out. The $\beta$- and $\alpha$-decays of these nuclei lead to the
stable nuclei $^{206}$Pb, $^{207}$Pb, $^{208}$Pb, and $^{209}$Bi, or the
long-lived actinides. Therefore, the progenitor yields at $r$-process
freeze-out are buried in the solar $r$-process abundances at $A=206$--209
and the solar abundances of the actinides $^{235}$U, 
$^{238}$U, $^{232}$Th and $^{244}$Pu. So far, theoretical calculations
(e.g., Pfeiffer, Kratz, \& Thielemann 1997; Cowan et al. 1999)
aiming to deduce the conditions (e.g., temperature, neutron number density,
and freeze-out time) required for production of nuclei beyond $^{209}$Bi
are based on these solar system data. However, the solar $r$-process
abundances at $A=206$--209 are subject to uncertainties in our understanding
of the $s$-process (e.g., 
production in AGB stars and contributions from
stars with different metallicities), and the solar abundances of the
actinides depend on their long-term 
production history as well as the yields of
their progenitors due to their decay after production. 
In addition, it has been noted that the solar $r$-process abundance pattern
may not represent a single kind of $r$-process events, 
but that the production sites of $r$-process nuclei
may differ below and above the abundance peak at $A\sim 130$ 
(Wasserburg, Busso, \& Gallino 1996).
By comparison,
gamma-ray signals associated with the decay of nuclei beyond $^{209}$Bi
can provide much better guidance for theoretical $r$-process calculations
as they probe the progenitor yields from individual $r$-process events.

Gamma-ray signatures of a supernova $r$-process event were discussed earlier 
by Meyer \& Howard (1991). They identified 
$^{125}$Sb, $^{137}$Cs, $^{144}$Ce, $^{155}$Eu, and $^{194}$Os
with lifetimes of $\sim 1$--40 yr as the
relevant nuclei and estimated the corresponding gamma-ray fluxes from
SN~1987A. Qian, Vogel, \& Wasserburg (1998b) generalized this approach
with particular consideration given to future gamma-ray detectors.
The latter workers found that a detector with a sensitivity of
$\sim 10^{-7}\ \gamma\ {\rm cm}^{-2}\ {\rm s}^{-1}$ at energies of
$\sim 100$--700~keV may detect gamma rays from the decay of: 
(1) the above nuclei produced in a future Galactic supernova; (2)
$^{126}$Sn (with a lifetime of $\sim 10^5$~yr)
in the Vela supernova remnant; and (3) $^{126}$Sn produced
by  past supernovae in the Galaxy. The scientific returns of such
detections would be establishment of supernovae as the $r$-process site
[(1) and (2)], test of distinct supernova sources for $r$-process nuclei
below and above $A\sim 130$ [(1), see Wasserburg et al 1996;
Qian, Vogel, \& Wasserburg 1998a], and elimination of neutron star-neutron
star mergers as the possible 
main source of $r$-process nuclei near $A\sim 126$ [(3)].

Gamma rays associated with the decay of nuclei beyond $^{209}$Bi in supernova
remnants were discussed first by Clayton \& Craddock (1965). They estimated
fluxes from the Crab nebula assuming the theoretical yields of 
Seeger, Fowler, \& Clayton (1965) normalized to $1.5\times 10^{-4}\ M_\odot$
of $^{254}$Cf. This enormous amount of 
production was motivated by the speculation
of Burbidge et al. (1957) that the early supernova light curves
are powered by the spontaneous fission of $^{254}$Cf. We note that 
Clayton \& Craddock (1965) pointed out the significance of detecting the
gamma rays from the decay of nuclei beyond $^{209}$Bi. These gamma rays are
the focus of this paper. They
were not discussed by Qian et al. (1998b) mainly because the combination of
expected low yields and lifetimes of $\sim 10^3$~yr for the relevant nuclei
requires a nearby supernova remnant much closer than Crab ($\sim 1$~kpc
away) or much younger than Vela ($\sim 10^4$~yr of age) to provide
gamma-ray fluxes of $\sim 10^{-7}\ \gamma\ {\rm cm}^{-2}\ {\rm s}^{-1}$.

Intriguingly, a new supernova remnant (RX~J0852.0--4622/GRO~J0852--4642) 
near Vela was discovered recently via its X-ray emission (Aschenbach 1998) 
and gamma rays from $^{44}$Ti decay (Iyudin et al. 1998). As $^{44}$Ti has
a lifetime of $\sim 90$~yr, the detection of 
the corresponding decay gamma rays clearly establishes this remnant
as a young object with an estimated age of $\sim 700$~yr.
Its distance is estimated to be $\sim 200$~pc.
These parameters are consistent with X-ray and gamma-ray observations
leading to the discovery and are adopted for our discussion here.
Although the nature of the supernova associated
with this remnant cannot be established directly by observations yet,
Chen \& Gehrels (1999) argued that the current expansion of the remnant
is too slow for its young age to
be due to a Type Ia supernova. In the following discussion, we assume that
it was produced by a core-collapse supernova.

With a distance of $\sim 200$~pc and an age of $\sim 700$~yr for the
new supernova remnant, we find that this remnant
could provide gamma-ray fluxes
of $\sim 10^{-7}\ \gamma\ {\rm cm}^{-2}\ {\rm s}^{-1}$ from the decay
of $^{226}$Ra, $^{229}$Th, $^{241}$Am, $^{243}$Am, $^{249}$Cf, and
$^{251}$Cf with lifetimes of $\sim 500$--$10^4$~yr. These fluxes are
estimated in \S2. We also find that 
for future supernovae that may occur at distances of $\sim 1$~kpc, 
the decay of $^{227}$Ac and $^{228}$Ra
with lifetimes of 31.4 and 8.30~yr, respectively, can produce fluxes of
$\sim 10^{-7}\ \gamma\ {\rm cm}^{-2}\ {\rm s}^{-1}$. 
This is presented in \S3. We further discuss the
significance of detecting the relevant
fluxes and give our conclusions in \S4. 

\section{Expected gamma-ray fluxes from the newly discovered supernova
remnant}

In this section, we estimate the gamma-ray fluxes due to the decay of
nuclei with $A>209$ in the newly discovered supernova remnant.
A supernova or its remnant becomes transparent to gamma rays $\sim 1$~yr 
after the explosion. The flux at a specific energy  $E_\gamma$ 
from the decay of a radioactive nucleus ``$i$'' 
produced in the supernova is 
\begin{eqnarray}
F_{\gamma,i}&=&{N_A\over 4\pi d^2}{(\delta M)_i\over A}
{I_{\gamma,i}\over\bar\tau_i}\exp\left(-{t\over\bar\tau_i}\right)\ 
=\ 7.9\times 10^{-7}\ \gamma~{\rm cm}^{-2}~{\rm s}^{-1}\nonumber\\
&&\times\,I_{\gamma,i}
\left[{(\delta M)_i\over 2\times 10^{-8}\,M_\odot}\right]
\left({200\over A_i}\right)\left({10^3\ {\rm yr}\over\bar\tau_i}\right)
\left({200\ {\rm pc}\over d}\right)^2 \exp\left(-{t\over\bar\tau_i}\right),
\label{flux}
\end{eqnarray}
where $N_A$ is Avogadro's number, $(\delta M)_i$ is the amount of production
(by mass) for this nucleus, $A_i$ is its mass number,
$\bar\tau_i$ is its lifetime, $I_{\gamma,i}$ is the number of photons
emitted at $E_\gamma$ per decay of this nucleus, and
$d$ and $t$ are the distance and 
the age of the supernova remnant, respectively.

From equation (\ref{flux}), we see that for comparable $I_{\gamma,i}$ and
$(\delta M)_i$, optimal fluxes are obtained for those nuclei with $\bar\tau_i$
comparable to the age $t\sim 700$~yr of the newly discovered 
supernova remnant.
Accordingly, we select $^{226}$Ra, $^{241}$Am, $^{249}$Cf, and $^{251}$Cf
as the nuclei of potential interest to future gamma-ray experiments.
The lifetimes of these nuclei range from
506~yr for $^{249}$Cf to $2.31\times 10^3$~yr for $^{226}$Ra (see Table 2).
For completeness, we also include $^{229}$Th and $^{243}$Am with somewhat
longer lifetimes of $1.06\times 10^4$~yr. The above nuclei were also
considered by Clayton \& Craddock (1965). 
For nuclei with substantially shorter lifetimes, 
any original abundances produced by chains of direct $\beta$-decays after
$r$-process freeze-out have already decayed away.
On the other hand, nuclei with
substantially longer lifetimes do not produce significant gamma-ray
fluxes due to their slow decay although they may be abundant 
in the new remnant.
Finally, fast decays of the daughter nuclei of 
our selected parent nucleus ``$i$''
are in equilibrium with the slow decay of their parent. The fluxes 
due to such fast decays are also given by equation (\ref{flux}) with
appropriate assignment of $I_{\gamma,i}$.

The key quantity for estimating $F_{\gamma,i}$ is $(\delta M)_i$, 
the amount of a radioactive nucleus ``$i$'' produced in the supernova 
associated with the new remnant. 
We first calculate the average amounts $\langle\delta M\rangle_i$ of the 
relevant nuclei produced in a supernova using the solar abundances of 
the actinides. We assume that the production of
such nuclei by past Galactic supernovae occurred at a uniform rate before
solar system formation (see Wasserburg et al. 1996).
Each of these nuclei are produced by a chain of direct $\beta$-decays 
after $r$-process freeze-out and by $\alpha$- and $\beta$-decays 
of progenitors from 
other chains (cf. Fowler \& Hoyle 1960). The contributions from 
possible progenitors with much heavier masses are limited by fission. 

With the above assumption, the time evolution of  
the total mass $M_{232}$ of $^{232}$Th in the Galaxy
is governed by
\begin{equation}
\dot M_{232}=\langle\delta M\rangle_{232}f_{\rm SN}
-{M_{232}\over\bar\tau_{232}},
\end{equation}
where $f_{\rm SN}$ is the Galactic supernova frequency 
and $\bar\tau_{232}$ is
the lifetime of $^{232}$Th. At the time of solar system formation, we have
\begin{equation}
M_{232}^{\rm SSF}=\langle\delta M\rangle_{232}f_{\rm SN}\bar\tau_{232}
[1-\exp(-T_{\rm UP}/\bar\tau_{232})],
\end{equation}
where $T_{\rm UP}$ is the duration of uniform production before solar system
formation. Further assuming that the solar composition is a Galactic 
average, we have
\begin{equation}
\langle\delta M\rangle_{232}={X_{\odot,232}^{\rm SSF}M_G\over 
f_{\rm SN}\bar\tau_{232}[1-\exp(-T_{\rm UP}/\bar\tau_{232})]},
\label{adm}
\end{equation}
where $X_{\odot,232}^{\rm SSF}$ is the solar mass fraction of $^{232}$Th
at solar system formation and $M_G$ is the total mass of stars in the Galaxy.
The above calculation has been repeated for the other actinides and
the results are given in Table 1 for $M_G=10^{11}~M_\odot$,
$f_{\rm SN}=(30~{\rm yr})^{-1}$, and $T_{\rm UP}=10^{10}$~yr.

From Table 1, we see that the calculated 
average amount of production per progenitor
for $^{232}$Th, $^{235}$U, and $^{238}$U, 
is $\langle\delta M\rangle/N_{\rm pro}\approx 2\times 10^{-8}\ M_\odot$. 
The value 
$\langle\delta M\rangle/N_{\rm pro}\approx 0.7\times 10^{-8}\ M_\odot$
for $^{244}$Pu is somewhat lower.
Among the nuclei of interest to us, $^{226}$Ra, $^{229}$Th,
$^{241}$Am, and $^{243}$Am each receive contribution from only one 
progenitor while $^{249}$Cf and $^{251}$Cf receive contributions from
three and two progenitors, respectively, $\sim 700$~yr after $r$-process
freeze-out. Taking 
$\langle\delta M\rangle/N_{\rm pro}\approx 2\times 10^{-8}\ M_\odot$
for nuclei below $^{244}$Pu and
$\langle\delta M\rangle/N_{\rm pro}\approx 0.7\times 10^{-8}\ M_\odot$
for those above $^{244}$Pu, we see that
$\langle\delta M\rangle_i\approx 2\times 10^{-8}~M_\odot$ seems to be
a reasonable estimate for the average amount of production in a supernova
for each of our selected nuclei.

To check how well $\langle\delta M\rangle_i$ represents the
actual amount of production $(\delta M)_i$ in a specific supernova,
we use the
observed Th abundances in very metal-poor stars in the Galactic halo.
Sneden et al. (1996, 1998) showed that the observed
abundances of elements including and beyond Ba ($A\sim 135$--195) 
in CS~22892--052 ($[{\rm Fe}/{\rm H}]\approx -3.1$),
HD~115444 ($[{\rm Fe}/{\rm H}]\approx -2.8$), and
HD~122563 ($[{\rm Fe}/{\rm H}]\approx -2.7$) 
follow the solar $r$-process abundance pattern 
remarkably closely. This strongly argues that the $r$-process
already occurred in the very early history of the Galaxy and favors its
association with supernovae, the progenitors of which 
evolve on timescales of $\sim 10^7$~yr. 
If the observed $r$-process
elemental abundances in a very metal-poor halo star reflect the
yields from a single supernova (see, e.g., Audouze \& Silk 1995), 
we can estimate the amount 
$(\delta M)_{232}$ of $^{232}$Th produced in this supernova as follows.

We picture that early in the Galactic history, a 
supernova exploded and the expansion of its debris was slowed down by
the interstellar medium (ISM). We assume that the debris and the ISM were
well mixed inside the supernova remnant where metal-poor halo stars 
were formed. With these assumptions, the amount of Th produced
in the supernova is
\begin{equation}
(\delta M)_{232}\approx 232\left({N_{232}\over N_H}\right)_\star M_H
\exp\left({t_\star\over\bar\tau_{232}}\right),
\end{equation}
where $M_H$ is the total mass of hydrogen inside the supernova remnant,
$t_\star$ is the age of the halo star, and  
\begin{equation}
\left({N_{232}\over N_H}\right)_\star=10^{\log\epsilon_{232,\star}-12}
\end{equation}
is the observed Th/H abundance ratio given in the usual spectroscopic
notation $\log\epsilon_{232,\star}$.
The supernova debris would mix with the ISM 
until its original energy/momentum was dispersed for the supernova remnant
to blend in with the general ISM.
Taking $M_H\approx 4\times 10^4~M_\odot$ as a typical value for supernova 
remnants in very metal-poor interstellar medium (e.g., Thornton et al. 1998)
and assuming $t_\star\approx 1.5\times 10^{10}$~yr, we obtain
\begin{equation}
(\delta M)_{232}\approx 1.9\times 10^{-5+\log\epsilon_{232,\star}}~M_\odot.
\label{dm}
\end{equation}

Cowan et al. (1999) gave $\log\epsilon_{232,\star}=-1.6$ and $-2.1$ for 
CS~22892--052 and HD~115444, respectively. From equation (\ref{dm}), the
amounts of Th produced in the corresponding supernovae are
$(\delta M)_{232}\approx 4.8\times 10^{-7}$ and $1.5\times 10^{-7}~M_\odot$.
These results are in remarkable accord with the average value 
$\langle\delta M\rangle_{232}\approx 9.2\times 10^{-8}~M_\odot$
found from the previous calculation (eq. [\ref{adm}]).
We conclude that the values given in Table 1 are good estimates for 
the actinide yields of an individual supernova. However, it is important 
to note that Sneden et al. (1998) also found a very metal-poor halo star 
(HD~122563) with essentially the same value of [Fe/H] as HD~115444, 
but in which $^{232}$Th 
was not detected and the abundances of the
other heavy $r$-process nuclei are lower 
by a factor of $\sim 10$. This nondetection result would only yield an 
upper bound for $(\delta M)_{232}$. In addition, this result cannot be 
explained easily if we assume that there was only a single kind of supernovae
producing $r$-process nuclei, unless Fe and the $r$-process nuclei
were separated in the explosion and the mixing of the supernova debris 
with the ISM was not uniform. For the present paper, 
we will assume that the nearly concordant results for $(\delta M)_{232}$
and $\langle\delta M\rangle_{232}$ referred to above are correct.

From the above discussion, we assume
$(\delta M)_i=2\times 10^{-8}~M_\odot$ for each of the nuclei
$^{226}$Ra, $^{229}$Th, $^{241}$Am, $^{243}$Am, $^{249}$Cf, and $^{251}$Cf
produced in the supernova associated with the new remnant.
A detailed correction for the number of 
progenitors will not greatly change the results. 
The expected gamma-ray fluxes from the new remnant
are given in Table 2 for $d=200$~pc and $t=700$~yr.
Nuclear data used in calculating these fluxes are taken from
Firestone et al. (1996) and Lederer et al. (1978, for K X-rays).
The energies of the K X-rays in Table 2 range from
75--90~keV for Bi to 104--127~keV for Cm. These X-rays arise as follows.
A low-lying excited state (with an excitation energy of a few hundred keV
or less) of a heavy daughter nucleus produced by the decay of its parent
quite often deexcites through ejection of a K-shell electron.
The refilling of this K-shell vacancy by an electron from outer shells
then produces the K X-rays.
Note that we give only the total K X-ray flux from a decay chain.
For a specific element, 
the relative intensities of its K X-ray branches with different energies
can be found in the Table of Isotopes (e.g., Lederer et al. 1978).

In Figure 1, we plot the
spectral distribution of the fluxes presented in Table 2 
assuming an idealized detector
with an energy resolution corresponding to a full width at half maximum
of ${\rm FWHM}=0.1E_\gamma$. The flux of a prominent
emission feature in this figure is $\sim 1/23$ the corresponding value of
$dF_\gamma/d(\log E_{\gamma,{\rm keV}})$. Note that the fluxes at
$E_\gamma=60$, 107 (Cm K X-rays), 388, and 609 keV
($\log E_{\gamma,{\rm keV}}=1.78$, 2.03, 2.59, and 2.78) all have values of
$F_\gamma\gtrsim 10^{-7}\ \gamma\ {\rm cm}^{-2}\ {\rm s}^{-1}$.
With an age of $\sim 700$~yr for the newly discovered supernova remnant,
these should be among the most prominent features of gamma-ray emission
from this remnant,
as all the short-lived radioactivities have become extinct. As indicated
in \S1 and discussed further in \S4, the possible additional gamma-ray
signals associated with long-lived radioactivities in the new remnant are
those from the decay of $^{126}$Sn (with a lifetime of $\sim 10^5$~yr).
However, the spectrum of the $^{126}$Sn decay lines (Qian et al. 1998b) can
be distinguished from that shown in Figure 1 by a detector with energy
resolutions similar to those assumed for this figure.

\section{Expected gamma-ray fluxes from a future Galactic supernova}

In this section, we estimate the gamma-ray fluxes due to the decay of nuclei
with $A>209$ produced in a future Galactic supernova. 
The relevant nuclei must have
lifetimes of $\bar\tau\gtrsim 1$~yr in order to have substantial abundances
when the supernova becomes transparent to gamma rays. 
In addition, equation (\ref{flux})
indicates that these nuclei must have $\bar\tau\lesssim 100$~yr
in order to provide fluxes of 
$F_\gamma\sim 10^{-7}\ \gamma~{\rm cm}^{-2}~{\rm s}^{-1}$ from a supernova
at a distance of $d\sim 1$~kpc. 
This is different from the case in \S2 as the only nuclei
of interest are those with relatively short lifetimes 
($1~{\rm yr}\lesssim\bar\tau\lesssim 100~{\rm yr}$) produced by chains of
direct $\beta$-decays after $r$-process freeze-out.
We find two such nuclei: $^{227}$Ac and
$^{228}$Ra with $\bar\tau = 31.4$ and 8.30 yr, respectively. 
The fluxes from the decay of these two nuclei are 
\begin{equation}
F_{\gamma,i} = 3.2\times 10^{-5}I_{\gamma,i}
\left[{(\delta M)_i\over 2\times 10^{-8}\,M_\odot}\right]
\left({200\over A_i}\right)\left({{\rm yr}\over\bar\tau_i}\right)
\left({{\rm kpc}\over d}\right)^2\exp\left(-{t\over\bar\tau_i}\right)
\ \gamma~{\rm cm}^{-2}~{\rm s}^{-1},
\label{flux1}
\end{equation}
where $t$ is the time since the supernova explosion (detection is possible at
$t\gtrsim 1$~yr). 

For the case of $^{227}$Ac ($\bar\tau = 31.4$~yr), 
as all of its unstable daughter nuclei have lifetimes less than or much
less than 0.1~yr, any initial inventory of these nuclei 
from direct $\beta$-decays of their respective $r$-process progenitors will
have decayed away after $\sim 1$~year and their decays will be in
equilibrium with the decay of $^{227}$Ac. 
Hence the fluxes from the decay chain of $^{227}$Ac are
represented by equation (\ref{flux1}) 
with appropriate assignment of $I_{\gamma,i}$.
However, for the case of $^{228}$Ra ($\bar\tau = 8.30$~yr),
its immediate daughter nucleus $^{228}$Ac decays quickly to $^{228}$Th
($\bar\tau = 2.76$~yr) 
that has a much longer lifetime than all of the subsequent
unstable daughter nuclei. Thus only the fluxes (emitted by $^{228}$Th) from
the decay of $^{228}$Ac are represented by equation (\ref{flux1}).
The emission by the daughter nuclei subsequent to $^{228}$Th
will start at zero, build up to a maximum, and then decrease essentially
following the decay of $^{228}$Ra. It turns out that fluxes emitted by
these nuclei reach half of the maximum values at $t\approx 1$~yr and
peak at $t=4.55$~yr. In addition, the peak fluxes coincide with 
the fluxes calculated from 
equation (\ref{flux1}) for $t=4.55$~yr.

With the same amount of production 
$(\delta M)_i\approx 2\times 10^{-8}\,M_\odot$ for $^{227}$Ac and $^{228}$Ra,
the corresponding peak fluxes from their decay have been calculated 
for a supernova at a distance of $d=1$~kpc (see Table~3).
The peak fluxes before the ($^{224}$Ra) entry in this
table are calculated from equation (\ref{flux1}) with $t=0$ and
are very close to the fluxes at $t\sim 1$~yr after the supernova explosion. 
As explained in the previous paragraph, 
the peak fluxes after the ($^{224}$Ra) entry coincide with the fluxes
calculated from equation (\ref{flux1}) for $t=4.55$~yr
and occur at such a time after the explosion.

In Figure 2, we plot the spectral distribution of the gamma-ray signals 
presented in Table 3 again assuming an idealized detector with
${\rm FWHM}=0.1E_\gamma$. As the emission by the daughter nuclei
after $^{228}$Th in the decay chain of $^{228}$Ra has a different time
evolution compared with that by the other nuclei in Table 3, 
the spectral distributions at two different times after the
supernova explosion are plotted:
$t\approx 1$~yr in Figure 2(a) and $t=4.55$~yr in Figure 2(b).
Note that the fluxes at
$E_\gamma=76$ (Bi K X-rays), 240, 338, 583, 931, and 2615 keV
($\log E_{\gamma,{\rm keV}}=1.88$, 2.38, 2.53, 2.77, 2.97, and 3.42) all have
values of $F_\gamma>2\times 10^{-7}\ \gamma\ {\rm cm}^{-2}\ {\rm s}^{-1}$.
In addition, the most prominent flux shifts from
$F_\gamma\approx 1.6\times 10^{-6}\ \gamma\ {\rm cm}^{-2}\ {\rm s}^{-1}$
at $E_\gamma=931$~keV in Figure 2(a) to
$F_\gamma\approx 9.2\times 10^{-7}\ \gamma\ {\rm cm}^{-2}\ {\rm s}^{-1}$
at $E_\gamma=240$~keV in Figure 2(b).
We note that Figure 2 is just an illustration of the gamma-ray spectra for
$r$-process nuclei with $A>209$ corresponding to the lines listed in Table 3.
As mentioned in \S1 and discussed further in \S4, we expect additional
gamma-ray signals, in particular those studied by Qian et al. (1998b),
from a future supernova. However, the additional signals
will not obscure the most prominent features at $E_\gamma=240$ and 931 keV
shown in Figure 2.

\section{Discussion and conclusion}

We have estimated gamma-ray fluxes due to the decay of nuclei with $A>209$
from a supernova or a supernova remnant assuming that the $r$-process
occurs in supernovae. We find that a gamma-ray detector
with a sensitivity of $\sim 10^{-7}\ \gamma~{\rm cm}^{-2}~{\rm s}^{-1}$
at energies of $\sim 40$~keV to $\sim 3$~MeV (see Tables 2 and 3) may detect
fluxes due to the decay of $^{226}$Ra, $^{229}$Th, $^{241}$Am, $^{243}$Am, 
$^{249}$Cf, and $^{251}$Cf in the newly discovered supernova remnant near
Vela. Such a detector may also detect fluxes due to the decay of 
$^{227}$Ac and $^{228}$Ra produced in a future supernova at a distance of 
$\sim 1$~kpc. To our knowledge, detectors with similar sensitivities
have been proposed by Kurfess (1994, ATHENA: the Advanced Telescope for 
High Energy Nuclear Astrophysics) and Boggs, Harrison, \& Prince (1999).
For consideration in developing these and other future detectors,
we note that the fluxes in Tables 2 and 3 are estimated for point sources.
A supernova can be regarded as a point source to good approximation.
However, the new supernova remnant has an angular size of
$\sim 1^\circ$. Consequently, our estimated fluxes in Table 2 are expected 
for detectors with angular resolutions of $\sim 1^\circ$.

Another important factor to consider in developing future detectors is
their energy resolutions. We have plotted the spectral distributions of
the expected gamma-ray signals from the newly discovered supernova remnant
and those from a future Galactic supernova in Figures 1 and 2, respectively, 
assuming an idealized detector with ${\rm FWHM}=0.1E_\gamma$. 
These gamma-ray spectra correspond to the lines listed in Tables 2 and 3.
With an age of $\sim 700$~yr for the newly discovered supernova remnant,
the spectrum shown in Figure 1 should be among the most prominent features
of gamma-ray emission at $E_\gamma\sim 60$--600 keV from this remnant,
as all the short-lived radioactivities
have become extinct. The possible additional signals from the decay of the
long-lived $^{126}$Sn have similar fluxes but a different spectral distribution
(Qian et al. 1998b). Even in the presence of these additional signals, 
we find that the prominent feature at $E_\gamma=60$ keV shown in Figure 1 
can be distinguished by a detector with energy resolutions similar to those
assumed for this figure. The identification of this feature can then facilitate
the extraction of the spectrum shown in Figure 1.
On the other hand, gamma-ray signals associated with
all radioactivities with lifetimes exceeding $\sim 1$~yr are expected from
a future supernova. The additional signals not shown in Figure 2 will not
present a problem for a detector with very good energy resolutions, as every
line will be distinguished. By examining the additional signals discussed
by Qian et al. (1998b), we find that the most prominent features at
$E_\gamma=240$ and 931 keV shown in Figure 2 can be recognized by a detector
with energy resolutions similar to those assumed for this figure. Furthermore,
we note that for the case of a future supernova at a distance of $\sim 1$~kpc, 
the most prominent flux at $t\sim 1$~yr after the explosion is
$F_\gamma\sim 1.6\times 10^{-6}\ \gamma\ {\rm cm}^{-2}\ {\rm s}^{-1}$
at $E_\gamma=931$~keV and that at $t=4.55$~yr after the explosion shifts to
$F_\gamma\sim 9.2\times 10^{-7}\ \gamma\ {\rm cm}^{-2}\ {\rm s}^{-1}$
at $E_\gamma=240$~keV. This time evolution of the spectral features can help
in their identification provided that the supernova will be observed 
several times over a period of $\sim 5$~yr after the explosion.

Detection of gamma-ray fluxes due to the decay of nuclei with $A>209$
from a supernova or a supernova remnant is the best proof for a supernova
$r$-process site as these nuclei are produced solely by the $r$-process.
Further, such a detection provides the most direct information on yields
of progenitor nuclei with $A>209$ at $r$-process freeze-out, which can
offer valuable guidance for theoretical studies. Finally, such a detection
also provides a direct means of comparing the $r$-process yields 
in a single supernova event with the solar $r$-process abundance pattern.
A supernova at a distance of $\sim 1$~kpc would make such a comparison 
possible for the mass region $A=125$--228 (the expected fluxes due to
the decay of $^{125}$Sb, $^{137}$Cs, $^{144}$Ce, $^{155}$Eu, and $^{194}$Os
are $\sim 10^{-5}\ \gamma~{\rm cm}^{-2}~{\rm s}^{-1}$, see Qian et al. 1998b.
If we attribute $^{106}$Ru to the $r$-process, the mass region for comparison
can be extended down to $A=106$.).
We also expect fluxes of $\sim 10^{-7}\ \gamma~{\rm cm}^{-2}~{\rm s}^{-1}$
due to the decay of $^{126}$Sn in the new supernova remnant as in the case
of Vela at a similar distance (Qian et al. 1998b). This would enable us 
to compare the $r$-process yields at $A=126$ and 226--251 
in a single supernova event
with the corresponding solar $r$-process abundance data. Such comparisons
can test decisively the diversity of supernova $r$-process events 
suggested by Wasserburg et al. (1996). We also note that the
yields at $A>209$ relative to those at $A\sim 135$--195 are crucial to the
nucleochronology based on the observed Th/Eu abundance ratios in very
metal-poor halo stars (Cowan  et al. 1997, 1999). 
To conclude, we strongly urge that detectors with sensitivities of 
$\sim 10^{-7}\ \gamma~{\rm cm}^{-2}~{\rm s}^{-1}$ at energies of
$\sim 40$~keV to $\sim 3$~MeV be developed with serious consideration
given to gamma rays from the decay of $r$-process nuclei.

\acknowledgments

We want to thank Steve Boggs, Ed Fenimore, Fiona Harrison, 
and Tom Prince for discussions.
We also would like to thank the referee, Al Cameron, for helpful comments.
This work was supported in part by the US Department of Energy under 
contract W-7405-ENG-36 and grant DE-FG03-88ER-40397, and
by NASA under grant NAG 5-4076, Division Contribution 8595(1013).
Y.-Z. Q. was supported by the J. Robert Oppenheimer
Fellowship at Los Alamos National Laboratory.

\clearpage

\figcaption{Spectral distribution of gamma-ray signals expected from the
newly discovered supernova remnant (cf. Table 2) for an idealized detector
with an energy resolution corresponding to a full width at half maximum
of ${\rm FWHM}=0.1E_\gamma$. The flux of a prominent emission
feature is $\sim 1/23$ the corresponding value of 
$dF_\gamma/d(\log E_{\gamma,{\rm keV}})$. Note that the fluxes at
$E_\gamma=60$, 107 (Cm K X-rays), 388, and 609 keV 
($\log E_{\gamma,{\rm keV}}=1.78$, 2.03, 2.59, and 2.78) all have values of
$F_\gamma\gtrsim 10^{-7}\ \gamma\ {\rm cm}^{-2}\ {\rm s}^{-1}$. See text
for detail.}

\figcaption{Spectral distribution of gamma-ray signals expected from a
future Galactic supernova at (a) $t\approx 1$~yr and (b) $t=4.55$~yr after
the explosion (cf. Table 3) for an idealized detector
with an energy resolution corresponding to a full width at half maximum
of ${\rm FWHM}=0.1E_\gamma$.
The flux of a prominent emission
feature is $\sim 1/23$ the corresponding value of 
$dF_\gamma/d(\log E_{\gamma,{\rm keV}})$. Note that the fluxes at
$E_\gamma=76$ (Bi K X-rays), 240, 338, 583, 931, and 2615 keV 
($\log E_{\gamma,{\rm keV}}=1.88$, 2.38, 2.53, 2.77, 2.97, and 3.42) all have
values of $F_\gamma>2\times 10^{-7}\ \gamma\ {\rm cm}^{-2}\ {\rm s}^{-1}$.
As the time increases from $t\approx 1$~yr in (a) to $t=4.55$~yr in (b),
the emission from the decay chain
of $^{277}$Ac and from $^{228}$Th in the decay chain of $^{228}$Ra decreases,
whereas that from the daughter nuclei after $^{228}$Th 
in the decay chain of $^{228}$Ra builds up to a maximum.
Consequently, the most prominent flux shifts from
$F_\gamma\approx 1.6\times 10^{-6}\ \gamma\ {\rm cm}^{-2}\ {\rm s}^{-1}$
at $E_\gamma=931$~keV in (a) to 
$F_\gamma\approx 9.2\times 10^{-7}\ \gamma\ {\rm cm}^{-2}\ {\rm s}^{-1}$
at $E_\gamma=240$~keV in (b). See text for detail.}

\clearpage

\begin{deluxetable}{lccccc}
\footnotesize
\tablecaption{Average Amounts of Actinides Produced in a Supernova}
\tablehead{
\colhead{} &
\colhead{$\bar\tau$\tablenotemark{a}} &
\colhead{$X_\odot^{\rm SSF}$\tablenotemark{b}} &
\colhead{$\langle\delta M\rangle$\tablenotemark{c}} &
\colhead{} & \colhead{$\langle\delta M\rangle/N_{\rm pro}$}\\
\colhead{actinide} &
\colhead{($10^9$ yr)} & \colhead{$(\times 10^{-10})$} &
\colhead{$(10^{-8}~M_\odot)$} & \colhead{$N_{\rm pro}$\tablenotemark{d}} &
\colhead{$(10^{-8}~M_\odot)$}
}
\startdata
$^{232}$Th & 20.3 & 2.42 & 9.2 & 5.8 & 1.6\nl
$^{235}$U & 1.02 & 0.335 & 9.9 & 6.0 & 1.6\nl
$^{238}$U & 6.45 & 1.07 & 6.3 & 3.1 & 2.0\nl
$^{244}$Pu & 0.117 & $7.46\times 10^{-3}$ & 1.9 & 2.8 & 0.69\nl
\enddata
\tablenotetext{a}{Lifetime.}
\tablenotetext{b}{Solar mass fraction at solar system formation
taken from Anders \& Grevesse (1989, Th and U) and Hudson et al. (1989, Pu).}
\tablenotetext{c}{Average amount produced in a supernova estimated from
equation (\protect\ref{adm}) and the like for $M_G=10^{11}~M_\odot$,
$f_{\rm SN}=(30~{\rm yr})^{-1}$, and $T_{\rm UP}=10^{10}$~yr.}
\tablenotetext{d}{Number of progenitors with loss
through fission taken into account.}
\end{deluxetable}

\clearpage

\begin{deluxetable}{lcccccc}
\footnotesize
\tablecaption{Expected Gamma-Ray Fluxes from the Newly Discovered 
Supernova Remnant}
\tablehead{
\colhead{$r$-process} & 
\colhead{$\bar\tau$\tablenotemark{b}} & 
\colhead{$E_\gamma$} & \colhead{} & 
\colhead{$F_\gamma$\tablenotemark{d}}\\
\colhead{nucleus\tablenotemark{a}} & 
\colhead{($10^3$ yr)} & \colhead{(keV)} & \colhead{$I_\gamma$} &
\colhead{$(10^{-7}\ \gamma~{\rm cm}^{-2}~{\rm s}^{-1})$}
}
\startdata
$^{226}{\rm Ra}$ & 2.31 & & & \nl
($^{214}$Bi) & & 242 & 0.075 & 0.17\nl
& & 295 & 0.185 & 0.42\nl
& & 352 & 0.358 & 0.80\nl
& & Bi K X-rays\tablenotemark{c} & 0.216 & 0.48\nl
($^{214}$Po) & & 609 & 0.448 & 1.0\nl
& & 768 & 0.048 & 0.11\nl
& & 1120 & 0.148 & 0.33\nl
& & 1238 & 0.059 & 0.13\nl
& & 1764 & 0.154 & 0.34\nl
& & 2204 & 0.049 & 0.11\nl
$^{229}{\rm Th}$ & 10.6 & & & \nl
($^{225}$Ra) & & Ra K X-rays\tablenotemark{c} & 0.336 & 0.21\nl
($^{225}$Ac) & & 40.0 & 0.300 & 0.18\nl
($^{213}$Po) & & 440  & 0.261 & 0.16\tablebreak
$^{241}{\rm Am}$ & 0.624 & & & \nl
($^{237}$Np) & & 59.5 & 0.359 & 1.2\nl
$^{243}{\rm Am}$ & 10.6 & & & \nl
($^{239}$Np) & & 74.7 & 0.682 & 0.39\nl
($^{239}$Pu) & & 106  & 0.272 & 0.16\nl
& & Pu K X-rays\tablenotemark{c} & 0.476 & 0.27\nl
$^{249}{\rm Cf}$ & 0.506 & & & \nl
($^{245}$Cm) & & 333 & 0.146 & 0.46\nl
& & 388 & 0.660 & 2.1\nl
& & Cm K X-rays\tablenotemark{c} & 0.080 & 0.25\nl
$^{251}{\rm Cf}$ & 1.30 & & & \nl
($^{247}$Cm) & & 177 & 0.177 & 0.50\nl
& & 227 & 0.063 & 0.18\nl 
& & Cm K X-rays\tablenotemark{c} & 0.436 & 1.2\nl
\enddata
\tablenotetext{a}{A nucleus without parentheses is a parent nucleus.
Nuclei in parentheses are daughter nuclei 
that emit gamma rays and K X-rays.}
\tablenotetext{b}{Lifetime of a parent nucleus.}
\tablenotetext{c}{These K X-rays are produced by atomic transitions of
the respective daughter nuclei (see text for explanation). Their energies
are 75--90 keV for Bi, 85--103 keV for Ra, 100--121 keV for Pu, and
104--127 keV for Cm. See, e.g., Lederer et al. (1978) for details of branches
with different energies.}
\tablenotetext{d}{Flux 
estimated from equation (\protect{\ref{flux}})
for $(\delta M)_i=2\times 10^{-8}~M_\odot$, $d=200$~pc, and $t=700$~yr.}
\end{deluxetable}

\clearpage

\begin{deluxetable}{lcccccc}
\footnotesize
\tablecaption{Expected Gamma-Ray Fluxes from a Future Galactic Supernova}
\tablehead{
\colhead{$r$-process} &
\colhead{$\bar\tau$\tablenotemark{b}} &
\colhead{$E_\gamma$} & \colhead{} &
\colhead{$F_\gamma$\tablenotemark{d}}\\
\colhead{nucleus\tablenotemark{a}} &
\colhead{(yr)} & \colhead{(keV)} & \colhead{$I_\gamma$} &
\colhead{$(10^{-7}\ \gamma~{\rm cm}^{-2}~{\rm s}^{-1})$}
}
\startdata
$^{227}$Ac & 31.4 & & & \nl
($^{223}$Ra) & & 50.1 & 0.079 & 0.70\nl
	     & & 79.7 & 0.019 & 0.17\nl
             & & 93.9 & 0.014 & 0.12\nl
             & & 236  & 0.121 & 1.1\nl
             & & 256  & 0.069 & 0.62\nl
             & & 286  & 0.015 & 0.14\nl
             & & 300  & 0.023 & 0.20\nl
             & & 305  & 0.012 & 0.11\nl
             & & 330  & 0.027 & 0.24\nl
($^{219}$Rn) & & 122  & 0.012 & 0.11\nl
             & & 144  & 0.032 & 0.29\nl
             & & 154  & 0.056 & 0.50\nl
             & & 269  & 0.137 & 1.2\nl
             & & 324  & 0.039 & 0.35\nl
             & & 338  & 0.028 & 0.25\nl
             & & 445  & 0.013 & 0.11\nl
($^{215}$Po) & & 271  & 0.108 & 0.96\nl
             & & 402  & 0.064 & 0.57\nl
($^{211}$Bi) & & 405  & 0.038 & 0.34\nl
             & & 427  & 0.018 & 0.16\nl
             & & 832  & 0.035 & 0.31\nl
($^{207}$Tl) & & 351  & 0.129 & 1.1\tablebreak
$^{228}$Ra & 8.30 & & & \nl
($^{228}$Th) & & 99.5 & 0.013 & 0.43\nl
	     & & 129  & 0.024 & 0.82\nl
             & & 209  & 0.039 & 1.3\nl
             & & 270  & 0.034 & 1.2\nl
             & & 328  & 0.030 & 0.99\nl
             & & 338  & 0.112 & 3.8\nl
             & & 409  & 0.019 & 0.65\nl
             & & 463  & 0.044 & 1.5\nl
             & & 755  & 0.010 & 0.34\nl
             & & 772  & 0.015 & 0.50\nl
             & & 795  & 0.043 & 1.5\nl
             & & 830  & 0.005 & 0.18\nl
             & & 836  & 0.017 & 0.56\nl
             & & 840  & 0.009 & 0.32\nl
             & & 911  & 0.266 & 8.9\nl
             & & 965  & 0.051 & 1.7\nl
             & & 969  & 0.162 & 5.4\nl
             & & 1588 & 0.033 & 1.1\nl
             & & 1631 & 0.016 & 0.54\tablebreak
($^{224}$Ra) & & 84.4 & 0.013 & 0.25\nl
($^{220}$Rn) & & 241  & 0.040 & 0.77\nl
($^{212}$Bi) & & 239  & 0.433 & 8.4\nl
             & & 300  & 0.033 & 0.64\nl
	     & & Bi K X-rays\tablenotemark{c} & 0.315 & 6.1\nl
($^{212}$Po) & & 727  & 0.066 & 1.3\nl
             & & 785  & 0.011 & 0.21\nl
             & & 1621 & 0.015 & 0.29\nl
($^{208}$Tl) & & 40.0 & 0.011 & 0.21\nl
($^{208}$Pb) & & 277  & 0.023 & 0.44\nl
             & & 511  & 0.081 & 1.6\nl
             & & 583  & 0.304 & 5.9\nl
             & & 861  & 0.045 & 0.87\nl
             & & 2615 & 0.356 & 6.9\nl
\enddata
\tablenotetext{a}{A nucleus without parentheses is a parent nucleus. 
Nuclei in parentheses are daughter nuclei 
that emit gamma rays and K X-rays.} 
\tablenotetext{b}{Lifetime of a parent nucleus.}
\tablenotetext{c}{These K X-rays are produced by atomic transitions of Bi.
Their energies are 75--90 keV. See, e.g., Lederer et al. (1978)
for details of branches with different energies.}
\tablenotetext{d}{Peak flux 
for $(\delta M)_i=2\times 10^{-8}~M_\odot$ and $d=1$~kpc.
The peak fluxes before the ($^{224}$Ra) entry are calculated 
from equation (\protect{\ref{flux1}}) with $t=0$ and are very close to
the fluxes at $t\sim 1$~yr after the supernova explosion. The peak fluxes
after the ($^{224}$Ra) entry coincide with the fluxes calculated 
from equation (\protect{\ref{flux1}}) for $t=4.55$~yr and occur 
at such a time after the explosion (see text for explanation).}
\end{deluxetable}


\clearpage

\begin{references}

\reference{}
Anders, E., \& Grevesse, N. 1989, \gca, 53, 197

\reference{}
Aschenbach, B. 1998, \nat, 396, 141

\reference{}
Audouze, J., \& Silk, J. 1995, \apjl, 451, L49

\reference{}
Boggs, S., Harrison, F., \& Prince, P. 1999, private communication

\reference{}
Burbidge, E. M., Burbidge, G. R., Fowler, W. A., \& Hoyle, F. 1957,
Rev. Mod. Phys., 29, 547

\reference{}
Cameron, A. G. W. 1957, \pasp, 69, 201

\reference{}
Chen, W., \& Gehrels, N. 1999, \apjl, 514, L103 

\reference{}
Clayton, D. D., \& Craddock, W. L. 1965, \apj, 142, 189

\reference{}
Cowan, J. J., McWilliam, A., Sneden, C., \& Burris, D. L. 1997,
\apj, 480, 246

\reference{}
Cowan, J. J., Pfeiffer, B., Kratz, K.-L., Thielemann, F.-K., Sneden, C.,
Burles, S., Tytler, D., \& Beers, T. C. 1999, \apj, 521, in press

\reference{}
Firestone, R. B., Shirley, V. S., Baglin, C. M., Chu, S. Y. F., \&
Zipkin, J. 1996, Table of Isotopes (8th ed.; New York: Wiley)

\reference{}
Fowler, W. A. \& Hoyle, F. 1960, Annals of Physics, 10, 280

\reference{}
Hudson, G. B., Kennedy, B. M., Podosek, F. A., \& Hohenberg, C. M. 1989,
in Proc. 19th Lunar and Planetary Science Conf., 
ed. G. Ryder \& V. L. Sharpton (Cambridge: Cambridge Univ. Press), 547.

\reference{}
Iyudin, A. F., et al. 1998, \nat, 396, 142 

\reference{}
Kurfess, J. D. 1994, ATHENA Mission Proposal, NASA New Mission Concepts in
Astrophysics

\reference{}
Lederer, C. M., et al. 1978, Table of Isotopes (7th ed.; New York: Wiley)

\reference{}
Meyer, B. S., \& Howard, W. M. 1991, in Supernovae, ed. S. E. Woosley
(New York: Springer), 630

\reference{}
Pfeiffer, B., Kratz,  K.-L., \& Thielemann, F.-K. 1997, Z. Phys. A, 357, 235

\reference{}
Qian, Y.-Z., Vogel, P., \& Wasserburg, G. J. 1998a, \apj, 494, 285

\reference{}
Qian, Y.-Z., Vogel, P., \& Wasserburg, G. J. 1998b, \apj, 506, 868

\reference{}
Seeger, P. A., Fowler, W. A., \& Clayton, D. D. 1965, \apjs, 11, 121

\reference{}
Sneden, C., Cowan, J. J., Burris, D. L., \& Truran, J. W. 1998,
\apj, 496, 235

\reference{}
Sneden, C., McWilliam, A., Preston, G. W., Cowan, J. J., Burris, D. L.,
\& Armosky, B. J. 1996, \apj, 467, 819

\reference{}
Thornton, K., Gaudlitz, M., Janka, H.-Th., \& Steinmetz, M. 1998,
\apj, 500, 95

\reference{}
Wasserburg, G. J., Busso, M., \& Gallino, R. 1996, \apj, 466, L109

\end{references}
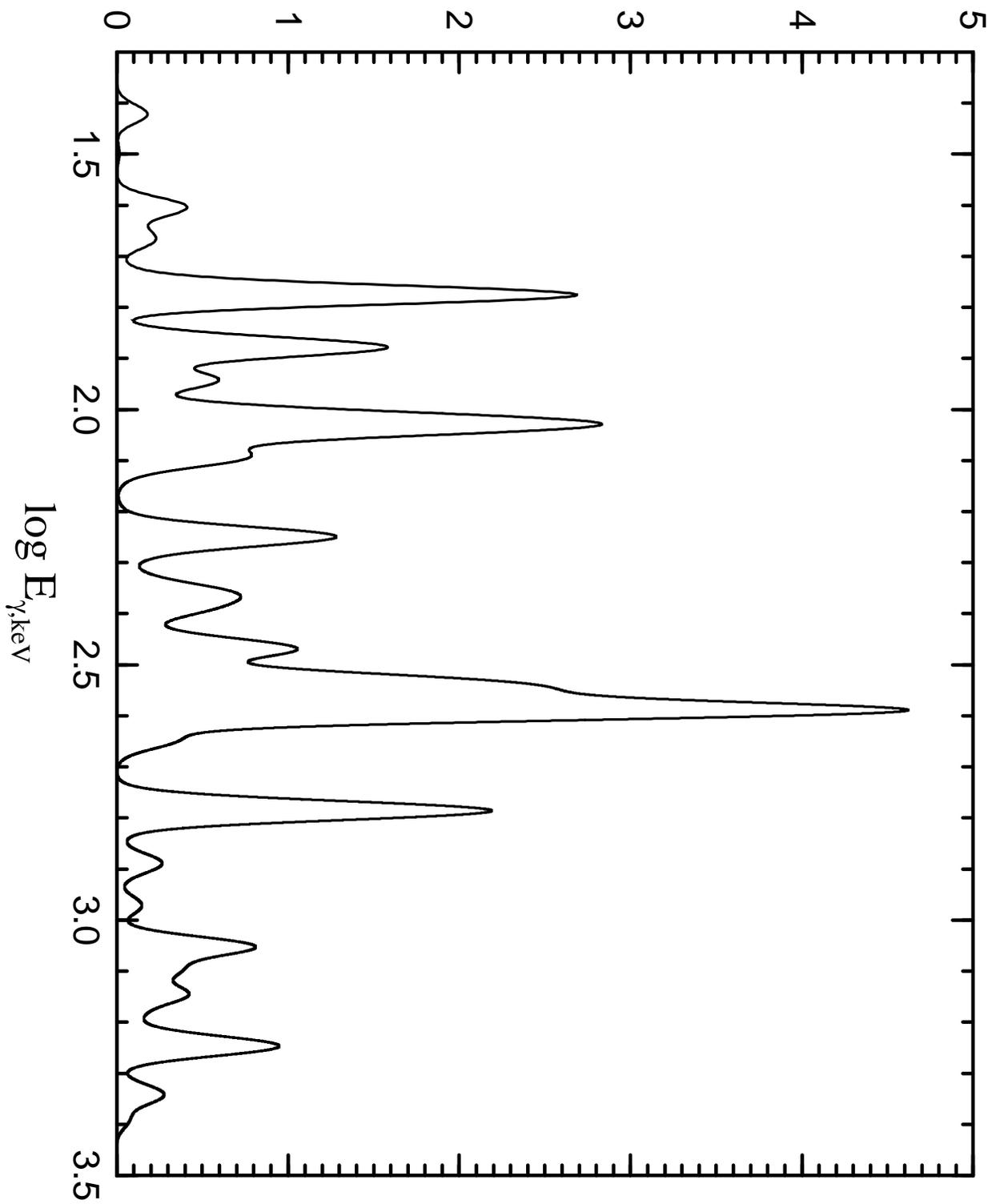
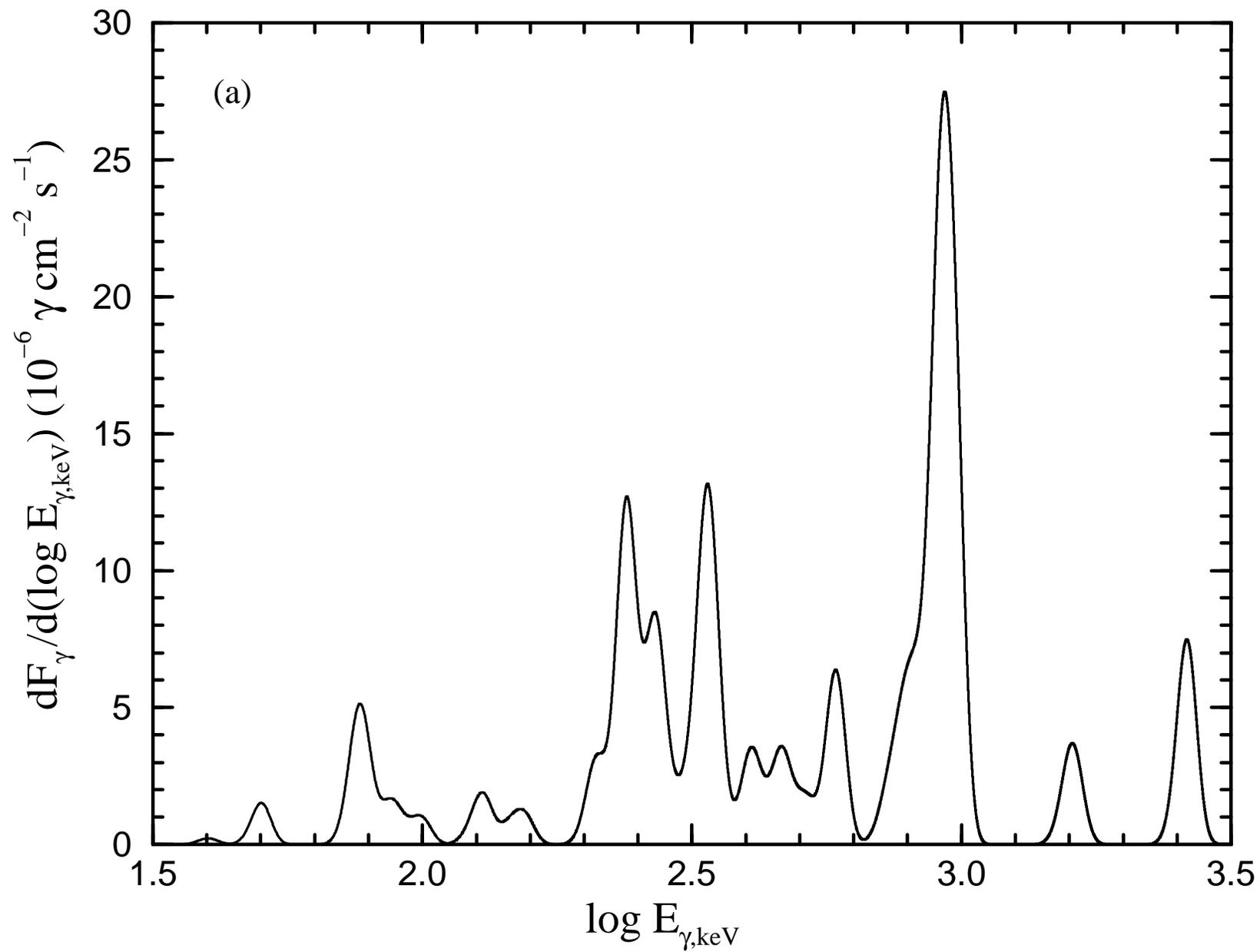
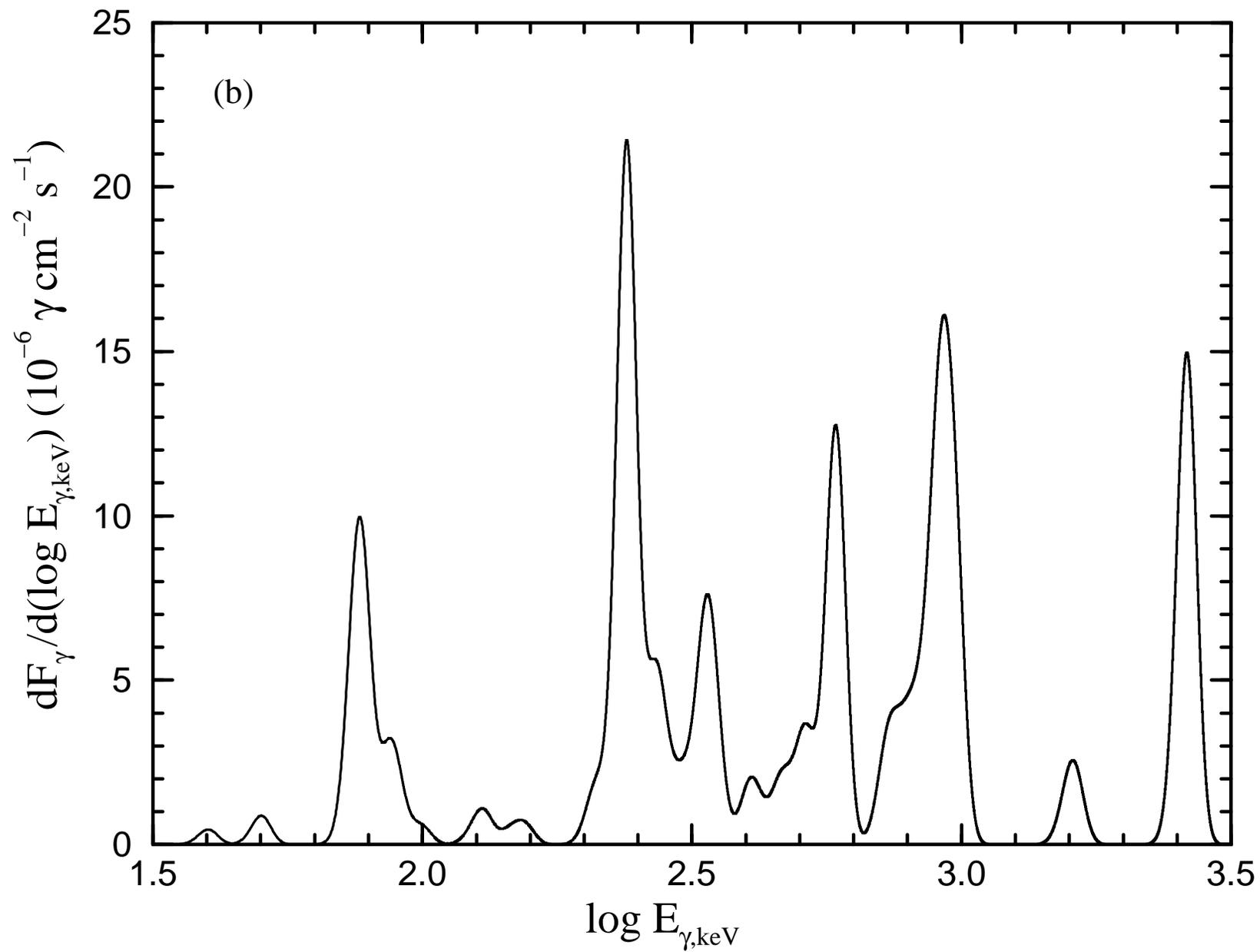
\end{document}